\documentclass[aps,showpacs,prb,superscriptaddress,floatfix,citeautoscript,reprint]{revtex4-1}
\usepackage{graphicx}
\usepackage{amsmath}
\usepackage{amssymb}
\usepackage{bm}
\usepackage{dcolumn}
\usepackage{subfigure}
\usepackage{units}
\usepackage{hyperref}
\usepackage[usenames]{color}
\usepackage{booktabs}
\usepackage{multirow}
\usepackage{siunitx}
\usepackage{wrapfig}
\usepackage{lipsum}
\usepackage{setspace}
\usepackage{microtype}
\usepackage{tikz}
\usepackage{comment}
\usepackage{xcolor}
\usepackage{float}
\hypersetup{pdfborder=0 0 0,colorlinks=true,citecolor=blue,linkcolor=blue}
\input epsf
\usepackage{xpatch}
\xpatchcmd\bibsection{19}{.3}{}{}

\begin{document}

\title{Collective plasmonic excitations in double-layer silicene at finite temperature}
\author{N. Dadkhah}
\affiliation{Department of Physics, Shahid Beheshti University, G. C., Evin, Tehran 1983969411, Iran}
\author{T. Vazifehshenas}
\email{t-vazifeh@sbu.ac.ir}
\affiliation{Department of Physics, Shahid Beheshti University, G. C., Evin, Tehran 1983969411, Iran}
\author{M. Farmanbar}
\affiliation{Faculty of Science and Technology and MESA+ Institute for Nanotechnology, University of Twente, P.O. Box 217, 7500 AE Enschede, The Netherlands}
\author{T. Salavati-fard}
\affiliation{Department of Physics and Astronomy, University of Delaware, Newark, DE 19716, USA}
\affiliation{Department of Chemical and Biomolecular Engineering, University of Houston, Houston, TX 77204, USA}

\begin{abstract}
We explore the temperature-dependent plasmonic modes of an n-doped double-layer silicene system which is composed of two spatially separated single layers of silicene with a distance large enough to prevent the interlayer electron tunneling. By applying an externally applied electric field, we numerically obtain the poles of the loss function within the so-called random phase approximation, to investigate the effects of temperature and geometry on the plasmon branches in three different regimes: topological insulator, valley-spin polarized metal, and band insulator. Also, we present the finite-temperature numerical results along with the zero-temperature analytical ones to support a discussion of the distinct effects of the external electric field and temperature on the plasmon dispersion. Our results show that at zero temperature both the acoustic and optical modes decrease by increasing the applied electric field and experience a discontinuity at the valley-spin polarized metal phase as the system transitions from a topological insulator to a band insulator. At finite temperature, the optical plasmons are damped around this discontinuity and the acoustic modes may exhibit a continuous transition. Moreover, while the optical branch of plasmons changes non-monotonically and noticeably with temperature, the acoustic branch dispersion displays a negligible growth with temperature for all phases of silicene. Furthermore, our finite-temperature results indicate that the dependency of two plasmonic branches on the interlayer separation is not affected by temperature at long wavelengths; the acoustic mode energy varies slightly with increasing the interlayer distance, whereas the optical mode remains unchanged.
\end{abstract}

\maketitle

\maketitle
\section{Introduction}
The advent of two-dimensional (2D) materials has sparked a considerable scientific interest due to their unique properties and their potential for applications in electronic, spintronic, valleytronic, optoelectronic and plasmonic devices. Silicene, the silicon counterpart of graphene has received much attention in recent years both theoretically\cite{guzmnverri2007,cahangirov2009,Enriquez2012,Kaltsas2013,LewYanVoon2016,Huang2013,barati2018} and experimentally\cite{kara2009,nakano2016,Sugiyama2010,lin2012,Feng2012,chiappe2012,Jamgotchian2012,Fleurence2012,Meng2013,Lalmi2010}. Like graphene, silicene is an atomically thin sheet of silicon atoms arranged in a honeycomb structure; however, unlike graphene, silicene does not hold a planar structure. A small out of plane buckling of the silicene structure due to the $sp^{3}$ hybridization results in a system of two sublattices (A and B) which lie in two parallel planes separated by a vertical distance $2l=0.46\mbox{\mbox{\mbox{\AA}}}$\cite{ni2012,drummond2012,cahangirov2009}. The novel properties of graphene arise from its band structure in which the valence and conduction bands meet at Dirac points ($K$ and $K^{\prime}$) in the first Brillouin zone with a well-defined linear energy dispersion relation around these points: $E(\mathbf{k})=\pm\hbar\nu_F|\mathbf{k}|$ with $ \nu_F=\unit[10^6]{m/s} $ and $\hbar\mathbf{k}$ being the carrier momentum. The strong spin-orbit interaction in silicene along with the buckled structure induces a nonzero band gap, $2\Delta_{SO}$, of about a few meV. The buckled structure of silicene provides new possibility to establish a potential difference between sublattices A and B by applying an external out-of-plane electric field, $E_{z}$. Interestingly, this feature allows tuning the band gap of silicene and discriminating between the different phases, topological insulator (TI), valley-spin polarized metal (VSPM) and band insulator (BI), which are clearly benefits over graphene. 
	
Furthermore, for the biased silicene-like materials, some valley-dependent properties can be expected in the presence of an exchange field. Having the extra spin and valley degrees of freedom, makes silicene very rich in behavior. Silicene can be effectively used as a source of nearly 100\% spin-polarized electrons\cite{Tsai2013} and as a valley-selective spin filter\cite{Zhai2016,wu2015jap,soodchomshom2014}. 
Moreover, silicene can probably be capable of being integrated into the current silicon-based electronics which makes it a promising material for future electronics applications\cite{tao2015}.

Considerable progress in fabrication techniques has promised synthesis of double-layer 2D structures. A double-layer structure is composed of two parallel electron or hole gas systems which are kept in a close vicinity and coupled through the Coulomb interaction. This class of 2D structures demand special attention due to exciting many-body phenomena, such as the coupled collective excitations, Coulomb drag, fractional quantized Hall effect\cite{Narozhny2016,Anorim2012,Zheng1993} and more, that they can display. The interlayer interaction accounts for unique effects in correlated double-layer systems which have been extensively investigated in double-layer GaAs-based quantum wells, double-layer graphene and double-layer phosphorene structures\cite{saberi2016,vazifehshenas2015,guven1997,gorbachev2012,hwang2009,peres2011,suen1992,kim2011,flensberg1995}.  

Plasmons, the collective excitations of an interacting electron or hole gas system, can be pictured as different modes of charge density oscillations. Plasmon branches are the central concepts of the rapidly developing field of plasmonics which deals with light-matter (photon-surface plasmon) interactions and their technologically significant consequences. The plasmonic dispersion relations which reveal the wave vector-dependent modes can be theoretically obtained from zeros of the dynamical dielectric function of the system\cite{sarma1981,vazifehshenas2010} and may be observed by inelastic light scattering spectroscopy, frequency domain far-infrared spectroscopy or inelastic electron scattering spectroscopy\cite{hwang2009}. The well-behaved plasmon modes are lying outside the single-particle excitation (SPE) region, however, those modes entering this region experience the Landau damping. 

The plasma oscillations of a single-layer silicene (SLS) have already been calculated both at zero temperature and finite temperature with and without a perpendicularly applied electric field\cite{wu2015,tabert2014,iurov2016,iurov2017}. It has been shown that, in an extrinsic SLS, the plasma oscillations follow a $\sqrt{q}$ behavior in the long-wavelength limit at zero temperature for all values of electric field (i.e.  all different phases). In addition, the location of the long-wavelength plasmon branch as a function of the applied electric field depends on the position of the Fermi energy with respect to the spin-split bands.\cite{tabert2014,chang2014} These results resemble those of single-layer graphene (SLG) and single-layer gapped graphene (SLGG) with $\Delta_{SO}=0$ and $\Delta_{SO}\ne0$, respectively. Moreover, the presence of a small gap in SLGG and various combinations of two gaps in SLS leads to the splitting of the plasmon branch and the appearance of new undamped plasmon modes.\cite{pyatkovskiy,iurov2016} On the other hand, in the case of SLG, plasmons decay into particle-hole pairs in the interband portion of the SPE region, while for SLS both intra- and interband dampings can be occurred by tuning the external electric field. \cite{tabert2014} 
The extrinsic (doped) finite-temperature plasmons of SLS have also been studied: as temperature increases from $T=0$, the plasmon energy first lowers, reaching to a minimum value and then bounces back in both cases of SLS and SLGG.\cite{iurov2017,patel2015}

Similar calculations were performed at zero temperature when both electric and magnetic (exchange) fields were applied\cite{peeters}. It is worth pointing out that an exchange field can be induced by positioning ferromagnetic adatoms \cite{Qiao2010} on the surface or by employing a ferromagnetic substrate \cite{Haugen2008,Yokoyama2014}. In the presence of these fields, the SPE region is given a spin-valley texture which results in the spin-polarized plasmons with zero $E_z$ but finite exchange field, and valley-spin-polarized plasmons when both fields are finite. However, it was mentioned that the magnetic field strongly affected the plasmon branch intensity such that it was unlikely to be experimentally observable. Also, the typical $\sqrt{q}$ behavior of long-wavelength plasmons in 2D systems was preserved when the magnetic field was turned on. For shorter wavelengths, although the plasmon dispersion looked different in the presence of a strong magnetic field, it was still possible to tune plasmons to display an approximate $\sqrt{q}$ behavior\cite{peeters}. In a recent work, the plasmon-phonon coupling in a valley-spin polarized SLS was investigated and the dispersion relations of hybrid modes were discussed\cite{Mirzaei2018}. 

Plasma oscillations have also been studied in a doped double-layer graphene both at zero \cite{stauber2012,Profumo2012,Gumbs2016,hwang2009,hwang2007,Sarma2009} and finite \cite{vazifehshenas2010,tuan2013,Badalyan2012a} temperatures within the so-called random phase approximation (RPA). At zero temperature and in the limit of small wave vectors, the plasmon dispersion of the in-phase (optical) branch obeys the $\sim \sqrt{q}$ relation and is independent of the interlayer separation, $d$. On the other hand, the out-of-phase (acoustic) branch has a $\sim q$ dispersion and depends upon layer distance as $\sim \sqrt{d} $. This long-wavelength behavior of plasmons is identical to that of the conventional double-layer 2D electron gas despite of different dependence on the electron density. Moreover, the plasmonic dispersion for larger wave vectors calculated within the RPA for both systems differs from one another.\cite{hwang2009,hwang2007,Sarma2009} At finite temperature, the frequency of the acoustic mode decreases down to $T \approx 0.4T_F$ and then increases afterwards. However, the slope of the growth/decline highly depends upon the value of the wave vector. The variation of optical mode energy with temperature is not uniform; it is fair to say that at low wave vectors, the optical branch mimics the behavior of the acoustic branch while at large wave vectors, the frequency of this branch generally increases with increasing temperature.\cite{vazifehshenas2010,tuan2013} 

In this paper, we consider an n-doped double-layer silicene (DLS) system in which two parallel single layers of silicene are placed in a dielectric environment at a distance $d$ from one another which is short enough to make an effective interaction between them but far enough to guarantee that there will be no electron tunneling. We investigate the effect of temperature on both plasmon branches with and without the presence of a perpendicular electric field within different regimes. We also study the dependence of the plasmon frequencies on the interlayer distance.

The rest of this paper is organized as follows: In Sec. \ref{sec:Theory}, we describe the model and theoretical formalism. In Sec. \ref{sec:results}, we present the results of calculations and discuss them in detail. Finally, a conclusion is given in Sec. \ref{sec:final}.  
\maketitle
\section{Theory}
\label{sec:Theory}
\subsection{Low-Energy Hamiltonian}
\label{sec:Low-energy Hamiltonian}
Focusing on low-energy phenomena, the effective Hamiltonian of a buckled honeycomb structure with an effective intrinsic spin-orbit coupling (SOC) strength $\Delta_{SO}$ ($ \Delta_{SO}=\unit[3.9]{meV} $ throughout this paper), in the presence of a perpendicular electric field, around the Dirac point $K_\xi$, is given by:\cite{kane2005,ezawaPRL,liu2011EffectiveHamiltonian}

\begin{equation}
\hat{H}_{\xi}=\hbar\nu_F(\xi k_x\hat{\tau_x}+k_y\hat{\tau_y})-\xi\Delta_{SO}\hat{\sigma}_z\hat{\tau}_z+\Delta_z\hat{\tau_z}
\label{Hamiltonian}
\end{equation}
Where $\xi$ distinguishes the two inequivalent valleys with $\xi=+1(-1)$ representing $K(K^{\prime})$ point. $\mathbf{k}=(k_x,k_y)$ is the 2D wave vector measured relative to the Dirac points. $\tau_i$ and $\sigma_i$ are Pauli matrices corresponding to (sublattice) pseudospin (written in A-B basis) and the real spin of the system, respectively. They are given as:
\begin{equation}
\tau_x=\begin{pmatrix}
0 & 1 \\
1 & 0
\end{pmatrix}
,
\tau_y=\begin{pmatrix}
0 & -i \\
i & 0
\end{pmatrix}
,
\tau_z=\sigma_z=\begin{pmatrix}
1 & 0 \\
0 & -1
\end{pmatrix}
\end{equation}
Further, $\nu_F$ is the Fermi velocity of the electrons, reported to be \unit[$ 5\times10^5 $]{m/s} for silicene\cite{ezawaNJP}. Finally $2\Delta_z$ shows the on-site potential difference between the two sublattices. This low-energy model works quite well for energies between \unit[-800]{} and \unit[+800]{meV}\cite{ezawaNJP}.

The effects of intrinsic Rashba SOC and electric field induced Rashba SOC are not considerable \cite{peeters} and are neglected in the Hamiltonian \ref{Hamiltonian}. Consequently, spin states are decoupled and the eigenvalues obtained from \ref{Hamiltonian} are labeled by valley ($ \xi $) and spin ($ \sigma $) indexes:
\begin{equation}
E_{\sigma\xi}(\mathbf{k})=\lambda \sqrt{\hbar^2 \nu_F^2 |\mathbf{k}|^2 +\Delta_{\sigma\xi}^2}
\label{dispersion}
\end{equation}
Where, $ \Delta_{\sigma\xi}=|\sigma \xi \Delta_{SO}-\Delta_z| $ and $ \lambda=\pm1 $ indexes over the conduction (particle) and valence (hole) bands, respectively. It is also helpful to introduce $ \Delta_{min}\equiv\Delta_{+1,+1}=\Delta_{-1,-1} $ and $ \Delta_{max}\equiv \Delta_{+1,-1}=\Delta_{+1,-1}$.
As it can be seen from the above equation, in the absence of an external electric field, there exists a band gap of $2\Delta_{SO}$. When $E_z$ is turned on and $ \Delta_z/{\Delta_{SO}}<1 $ (TI regime), the bands become spin-split, representing two energy gaps determined by $2\Delta_{min}$ and $2\Delta_{max}$. At $ \Delta_z/{\Delta_{SO}}=1 $ (VSPM regime), the lower band gap closes, however, by further increase of $ E_z $, it opens up again (BI regime).
\subsection{Collective Excitations}
\label{sec:B}
For a double-layer system, the location of the undamped plasmon collective modes, in the $(\omega,q)$ space, is determined by the zeros of the dielectric tensor, generalized from the single-layer dynamical dielectric function \cite{sarma1981}. 
For systems with high electron density it is reasonable to employ the RPA to calculate the dielectric tensor \cite{vazifehshenas2010}. In this approach, the aforementioned tensor is given by \cite{sarma1981,hwang2009}
\vspace{2mm}
\begin{equation}
\epsilon_{ij}(\omega,\mathbf{q})=\delta_{ij}-\nu_{ij}(\mathbf{q})\Pi^0_j(\omega,\mathbf{q}),
\label{DielectricTensor}
\vspace{2mm}
\end{equation}
where $ i $ and $ j $ specify the two layers, $ \nu_{ij}(\mathbf{q}) $ is the Fourier component of the bare 2D Coulomb interaction between electrons in layers $i$ and $j$ and $ \Pi^0_j $ is the non-interacting dynamical polarization function associated with $ j $th layer. For a double-layer system consisting of two identical layers separated by a distance $d$ we have $ \nu_{11}(q)=\nu_{22}(q)=\nu(q)=2\pi\alpha/q $ (intralayer interaction) and $ \nu_{12}(q)=\nu_{21}(q)=\nu(q)e^{-qd} $ (interlayer interaction) \cite{sarma1981} in which $ \alpha $ is the effective fine structure constant, expressed as $ e^2/{4\pi\epsilon_0\kappa} $ with $ \kappa $ being the effective background dielectric constant (throughout this paper we set $ \kappa=4 $ which corresponds to $\alpha/\hbar \nu_F=1.0944$). Finally, $ \Pi^0_i=\Pi^0_j=\Pi^0 $ is given by\cite{pyatkovskiy}:
\begin{multline}
\vspace{5mm}
	\Pi^0(\omega,q)=\frac{1}{8\pi ^ 2}\sum_{\sigma,\xi=\pm 1}^{}\int{d^2\mathbf{k}} \\
	\sum_{\lambda,{\lambda}^{\prime} =\pm 1}^{}(1+\lambda\lambda^{\prime}\frac{{\hbar^2 \nu_F^2}\;{\mathbf{k}.(\mathbf{q}+\mathbf{k})}+\Delta_{\sigma\xi}^2}{E_{\sigma\xi}(\mathbf{k}) E_{\sigma\xi}(\mathbf{k}+\mathbf{q})}) \\
	\times\frac{n_F(\lambda E_{\sigma\xi}(k))-n_F(\lambda^{\prime}E_{\sigma\xi}(\mathbf{k}+\mathbf{q}))}{\lambda E_{\sigma\xi}(\mathbf{k}) - \lambda^{\prime}E_{\sigma\xi}(\mathbf{k}+\mathbf{q})-\omega - i0^+}.
	\label{NIP}
\end{multline}
Here, $n_F(E)=1/[e^{((E-\mu)/{k_{B}T})}+1]$ is the Fermi-Dirac distribution function with $\mu$ being the chemical potential which equals to the Fermi energy $ E_F $ at $ T=0 $. It is important to notice that the upper limit of integration over $k$ is restricted by the low-energy condition.
Hence, in the case of identical layers, the plasmon modes can be obtained from the zeros of following relation
\begin{equation}
	\label{DiDeter}
	\epsilon(\omega,q)={(1-\nu(q)\Pi^0(\omega,q))}^2-\nu^2(q)e^{-2qd}\Pi^{0^2}(\omega,q).
\end{equation} 

The dynamical dielectric function is generally a complex function of $ (\omega,q) $ such that $ \omega $ explicitly has a positive imaginary part which guarantees the dynamical physical functions to be retarded. For undamped plasmon excitations in the $ (\omega,q) $ space, both real ($ \Re{\epsilon}(\omega,q)) $ and imaginary ($ \Im{\epsilon}(\omega,q) $) parts of dielectric function are equal to zero. However, there can be points where $ \Re{\epsilon(\omega,q)}=0 $ but $ \Im{\epsilon(\omega,q)} \neq 0 $ which indicates the existence of collective modes at these points but with a finite lifetime due to the nonzero value of $ \Im{\epsilon} $. These types of plasmons decay into particle-hole excitation states. For those points of the $ (\omega,q) $ plane where $ \Im{\epsilon(\omega,q)} \neq 0 $, the single-particle excitations (SPE) are possible, i.e. a photon with a wave vector $ q $ and frequency $ \omega $ can excite a particle-hole pair in the system.
These regions in the $ (\omega,q) $ space are called SPE spectrum; any collective mode presents in SPE will eventually decay (known as the Landau damping). SPE spectrum can be studied as two separate regions corresponding to the inter- and intra-band pair-creation excitations. 

At $ T=0 $ and in the limit of low energies and momenta, $\hbar\nu_Fq \ll \hbar\omega\ll E_F $ the non-interacting polarization function of Eq. (\ref{NIP}) has the following analytical form:\cite{pyatkovskiy,chang2014,peeters}

\begin{equation}
\label{ApproxPol}
\Pi^0(\omega,q) \approx \sum_{\xi,\sigma=\pm 1} \frac{q^2 E_F}{4\pi\hbar^2\omega^2}(1-{\Delta_{\xi\sigma}^2}/{E_F^2})\Theta(E_F-\Delta_{\xi\sigma}).
\vspace{5mm}
\end{equation}

For $ qd\ll 1 $, Eqs. (\ref{DiDeter}) and (\ref{ApproxPol}) give the corresponding plasmon frequencies as:
\begin{align}
\label{ApproxBrach+}
\tilde{\omega}_p^{OP} &\approx \sqrt{\tilde{\alpha}\tilde{q}\sum_{\sigma,\xi=\pm 1}^{}{(1-{\tilde{\Delta}_{\xi\sigma}^2})\Theta(1-\tilde{\Delta}_{\xi\sigma}})}  \\
\tilde{\omega}_p^{AP} &\approx \tilde{q} \sqrt{{\frac{1}{2}}{\tilde{\alpha}}{\tilde{d}}\sum_{\sigma,\xi=\pm 1}^{}{(1-{\tilde{\Delta}_{\xi\sigma}^2})\Theta(1-\tilde{\Delta}_{\xi\sigma}})},
\label{ApproxBrach-}
\end{align}
where $ \tilde{\alpha}=\alpha/{\hbar \nu_F}$, $ \tilde{d}=({E_F/{\hbar \nu_F}})d$, $ \tilde{\omega}={\hbar\omega}/E_F $, $ \tilde{q}={\hbar \nu_Fq}/E_F $, $ \tilde{\Delta}_{\xi\sigma}={\Delta_{\xi\sigma}}/E_F $ are all dimensionless parameters.
\begin{figure}[htp]
	\centering
	\includegraphics[width=8cm]{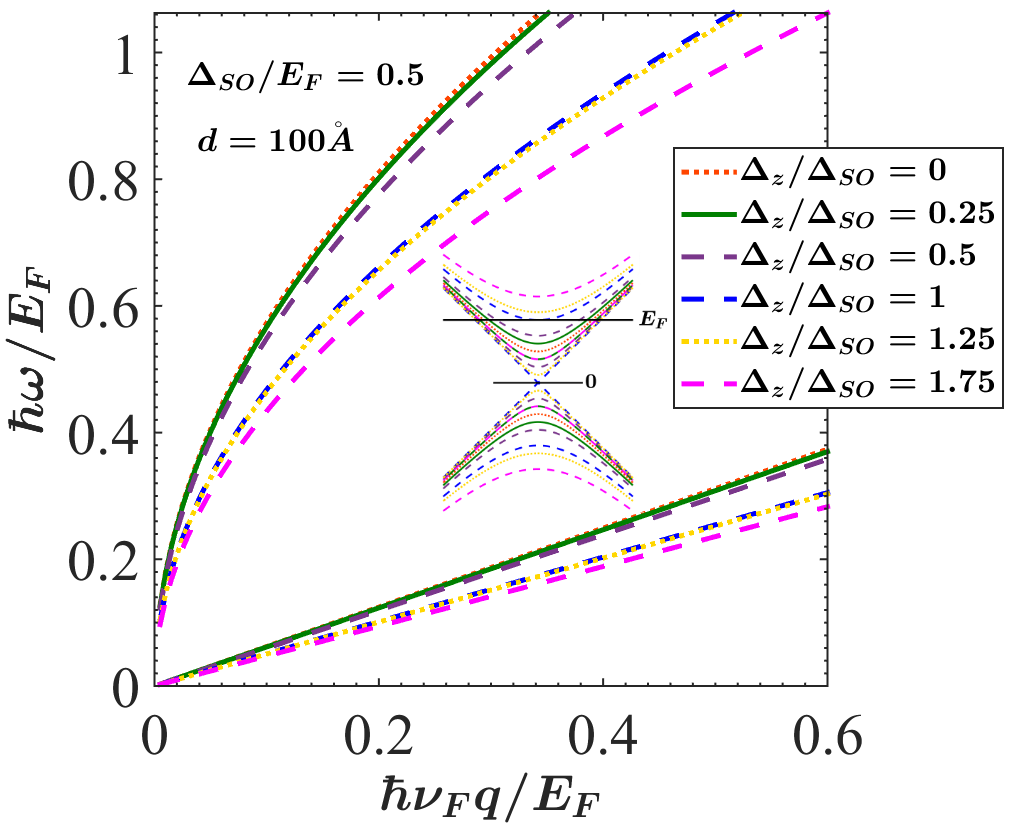}
	\caption{Zero-temperature plasmon dispersion of a DLS with  $ d=\unit[100]{\AA} $ for $ \Delta_{SO}/E_F=0.5 $ and varying $ \Delta_z $ at low energies. Inset shows the position of band structure relative to $ E_F $ as $ \Delta_z $ varies.} 
	\label{zeroPlasmon}
\end{figure}
\begin{figure}[htp]
	\centering
	\includegraphics[width=8cm]{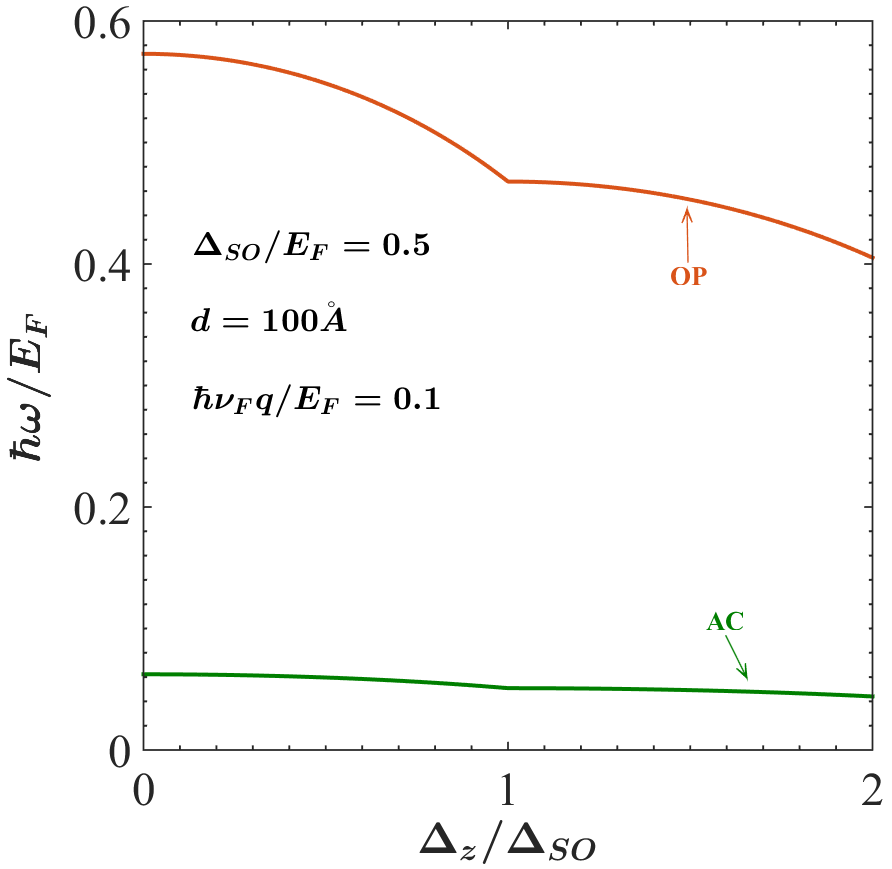}
	\caption{Zero-temperature plasmon frequency of a DLS as a function of $ \Delta_z $ with  $ d=\unit[100]{\AA} $ for $ \Delta_{SO}/E_F=0.5$ and $ \tilde{q}=0.1 $ at low energies.} 
	\label{zeroPlasmon_DeltaZ}
\end{figure}

Eqs. (\ref{ApproxBrach+}) and (\ref{ApproxBrach-}) reproduce the well-known behavior of plasmons in a double-layer system \cite{hwang2009}
which is shown in Fig. \ref{zeroPlasmon} for $ \Delta_{SO}/E_F=0.5 $,  $ d=\unit[100]{\AA} $ and varying $ \Delta_z $. The linear (non-linear) branch corresponds to the acoustic (optical) plasmon modes. Here, the Fermi energy sits above both gaps in the TI regime ($ \Delta_z/\Delta_{SO}<1 $) and only above the lower gap in the BI regime ($ \Delta_z/\Delta_{SO}>1 $). For the latter, Eqs. (\ref{ApproxBrach+}) and (\ref{ApproxBrach-}) show that the plasmon dispersion depends only on the lower band gap and the upper band gap plays no role. For this given configuration, the frequencies of both branches decrease as the potential difference increases. However, Fig. \ref{zeroPlasmon_DeltaZ} shows a discontinuity in the dispersion relations of both branches (which is more pronounced for the optical branch) as the system transitions from TI to BI through the VSPM phase ($ \Delta_z/\Delta_{SO}=1 $).

\section{Results and discussion}
\label{sec:results}
The spectral weight associated to a particular plasmon mode is directly proportional to $ \Im{(\epsilon^{-1}(\omega,q))} $ and is known as the loss function. This quantity which can be measured experimentally (e.g. in inelastic electron scattering experiments), gives a measure of the absorption of radiation and is an indication of the plasmon creation in the system. Thus, the loss function in the $ (\omega,q) $ space provides us with branches (poles of the loss function), which identifies both damped and undamped plasmons. In fact, the long-lived plasmons (those are outside the SPE regions), are $ \delta $-function peaks of the loss function, while the damped plasmons (those are inside SPE regions), correspond to the broadened peaks of the loss function \cite{hwang2009}.

\begin{figure}[htp]
	\includegraphics[width=8cm]{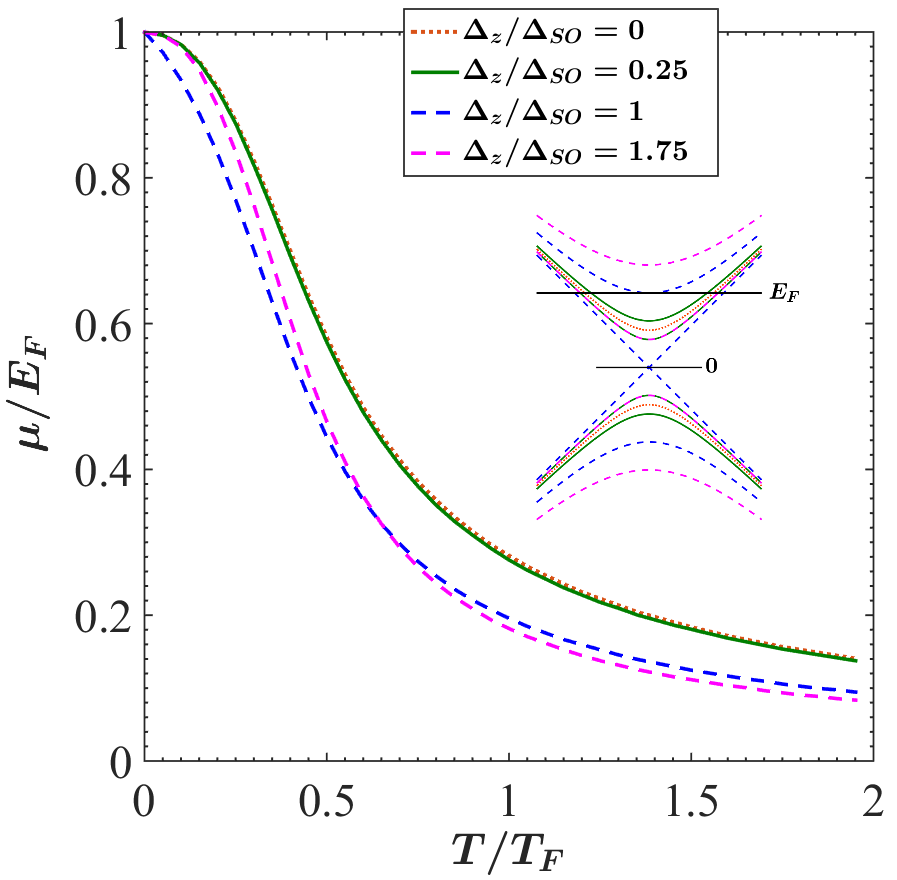}
	\caption{Scaled chemical potential as a function of dimensionless temperature. As expected, an increase in temperature results in decline of the chemical potential.}
	\label{muT}
\end{figure}

In this section, we first study the effect of temperature and band gap variations on the SPE regions. In this respect, we need to numerically calculate Eq. (\ref{NIP}). Note that, for $ T>0 $, the effect of temperature on the chemical potential, $ \mu $, must be taken into account. This can be done by assuming that the number of carriers of the system is fixed, and $ \mu $ changes, accordingly. Fig. \ref{muT} demonstrates the dependence of $ \mu/{E_F} $ on $ T/{T_F} $ calculated by using numerical methods for $ \Delta_{SO}/{E_F}=0.5 $ and different values of $ {\Delta_z}/{ \Delta_{SO}} $.

\begin{figure}[htp]
	\centering
	\includegraphics[width=8cm]{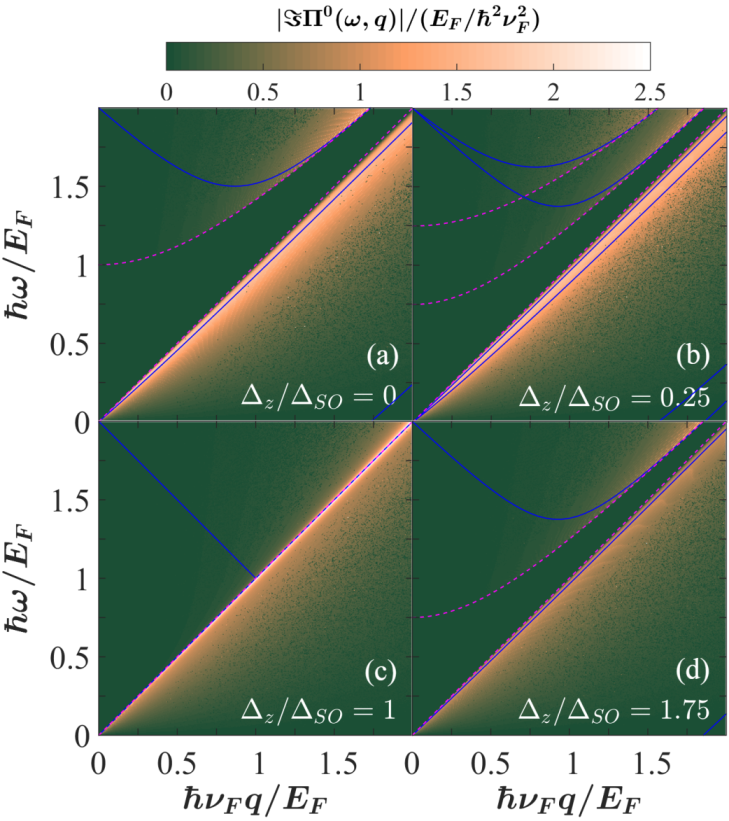}
	\caption{$ \Im \Pi^0(\omega,q) $, scaled by $ E_F/{{(\hbar\nu_F)}^2} $ for $ \Delta_{SO}/E_F=0.5 $ and varying $ \Delta_z/\Delta_{SO}:\;(a)\,0,\, (b)\,0.25,\, (c)\,1$ and $(d)\,1.75 $ at $ T/T_F=0.5$. Solid blue curves show the SPE boundaries at $ T=0 $ and dotted magenta curves are those of Eq. (\ref{Boundary}).}
	\label{SPEs_0.5}
\end{figure}

\begin{figure}[htp]
	\centering
	\includegraphics[width=8cm]{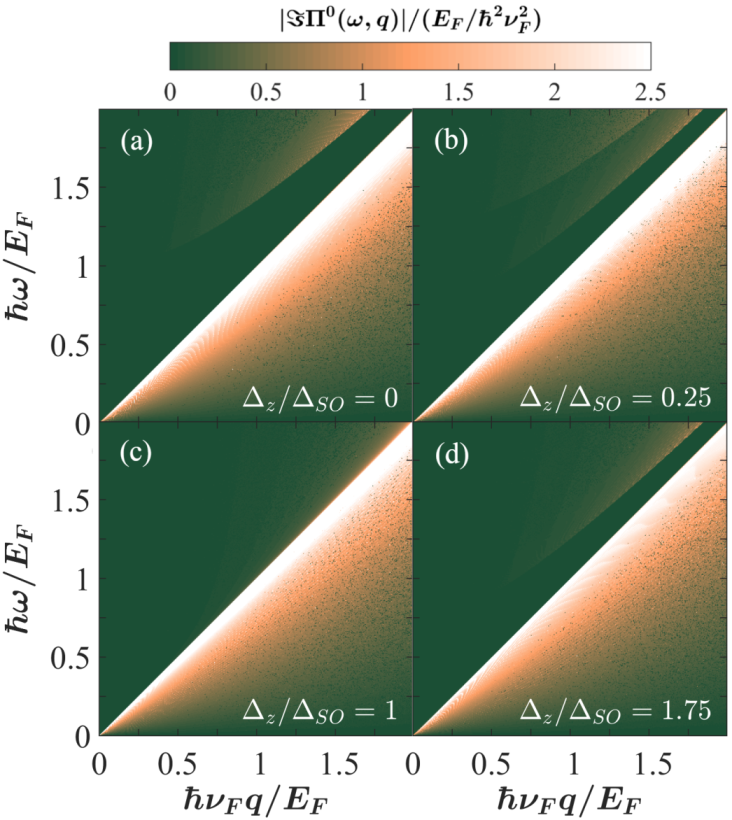}
	\caption{$ \Im \Pi^0(\omega,q) $, scaled by $ E_F/{{(\hbar\nu_F)}^2} $ for $ \Delta_{SO}/E_F=0.5 $ and varying $ \Delta_z/\Delta_{SO}:\;(a)\,0,\, (b)\,0.25,\, (c)\,1$ and $(d)\,1.75 $ at $ T/T_F=1$.}
	\label{SPEs_1}
\end{figure}

To investigate the SPE regions in silicene, we plot the imaginary part of the dynamical polarization function for $ \Delta_{SO}/{E_F}=0.5 $ at $ T=0.5T_F $ and $ T=T_F $, respectively (see Figs. \ref{SPEs_0.5} and \ref{SPEs_1}). The layers are apart by $ d=\unit[100]{\AA} $ (this value is used for the rest of calculations, as well) and the applied electric field varies $ \Delta_z/\Delta_{SO}=0, 0.25, 1$ and $1.75 $.  

The SPE$ ^{Inter} $ and SPE$ ^{Intra} $ regions are above and below the $ \tilde{\omega}=\tilde{q} $ line, respectively. The SPE boundaries are determined analytically at $ T=0 $ \cite{tabert2014}, whereas at finite temperature, they can be expressed approximately as:
\begin{multline}
\;\;\;\;\tilde{\omega}_{min(max)}^{th (Inter)}\approx2 \sqrt{(\tilde{q}/2)^2+(\tilde{\Delta}_{min(max)})^2}  \\
\tilde{\omega}_{min}^{Intra}\approx0 \\
\tilde{\omega}_{max}^{Intra}\approx\tilde{q} . \\
\label{Boundary}
\end{multline}
At a given $ \tilde{q} $, $ \tilde{\omega}^{th (Inter)} $ determines the threshold energy for onset of an interband  pair excitation, while $ \tilde{\omega}_{max}^{Intra} $ ($ \tilde{\omega}_{min}^{Intra} $) gives us the maximum (minimum) allowed energy transfer in an intraband transition. The zero- and finite-temperature SPE boundaries are displayed in Fig. \ref{SPEs_0.5} by the solid blue and dotted magenta curves, respectively for four different values of applied electric field.  As Figs. \ref{SPEs_0.5}\textcolor{blue}{(a)} and \ref{SPEs_1}\textcolor{blue}{(a)} show at $ \Delta_z=0 $ the SPE region is gapped and formation of undamped plasmons is possible in the system. When $ \Delta_z/\Delta_{SO}>0 $ (Figs. \ref{SPEs_0.5}\textcolor{blue}{(b-d)} and Figs. \ref{SPEs_1}\textcolor{blue}{(b-d)}), the band structure is spin-split and polarization function can be interpreted as the summation of two gapped subsystems with band gaps $ 2 \Delta_{min} $ and $ 2 \Delta_{max} $ \cite{tabert2014}. So, similar to $ T=0 $ case \cite{tabert2014,peeters}, SPE$ ^{Inter} $ region is divided into two parts such that the lower (upper) boundary corresponds to $ \Delta_{min} (\Delta_{max}) $. With increasing $ \Delta_z $, SPE$^{Inter}_{min} $ boundary moves toward the $ \tilde{\omega}=\tilde{q} $ line while the other one moves away. Furthermore, the intraband boundary, which is located slightly below the $ \tilde{\omega}=\tilde{q} $ line, moves toward the $ \tilde{\omega}=\tilde{q} $ line very slowly as $ \Delta_z $ increases. Thus the SPE gap becomes smaller. This pattern continues until $ \Delta_z=\Delta_{SO} $, where the lower boundary of SPE$^{Inter} $ and the SPE$^{Intra} $ boundary meet approximately at the $ \tilde{\omega}=\tilde{q} $ line resulting in closure of the SPE gap as it is shown in Figs. \ref{SPEs_0.5}\textcolor{blue}{(c)} and \ref{SPEs_1}\textcolor{blue}{(c)}. Consequently, at $ \Delta_z=\Delta_{SO} $ the possibility of undamped plasmon formation is negligible (except for small $ q $, $ \omega $ and $ T $ as it will be shown later). Fiqs. \ref{SPEs_0.5}\textcolor{blue}{(d)} and \ref{SPEs_1}\textcolor{blue}{(d)} show that as $ \Delta_z $ grows beyond the $ \Delta_{SO} $, the SPE gap opens again: the SPE$^{Inter} $ and SPE$^{Intra} $ boundaries both move away from the $ \tilde{\omega}=\tilde{q} $ line. 
Comparing Figs. \ref{SPEs_0.5} and \ref{SPEs_1} suggests that when the temperature increases from $ T=0.5T_F $ to $ T=T_F $, similar to the case of SLGG at finite temperature \citep{patel2015}, the boundaries are changed only slightly such that the extent of SPE gap decreases by temperature very slowly; this can be explained by the fact that single-particle transitions are more probable at higher temperatures\cite{patel2015}.
Moreover, in panels \textcolor{blue}{(a,b,d)} of Figs. \ref{SPEs_0.5} and \ref{SPEs_1}, the interband boundaries are not discernible before $ \tilde{q}\simeq0.25 $ but for values after that, that seems to be consistent with Eq. (\ref{Boundary}) and the intraband SPE boundary shows consistency with this equation, as well. For Panels \textcolor{blue}{c} where $ \Delta_z/\Delta_{SO}=1 $, Eq. (\ref{Boundary}) implies that the interband ($ \Delta_{min} $) and intraband boundaries lie along $ \tilde{\omega}=\tilde{q} $ line suggesting there is no SPE gap available for long-lived plasmons. This point will be discussed later on.

\begin{figure}[ht!]
	\centering
	\includegraphics[width=8cm]{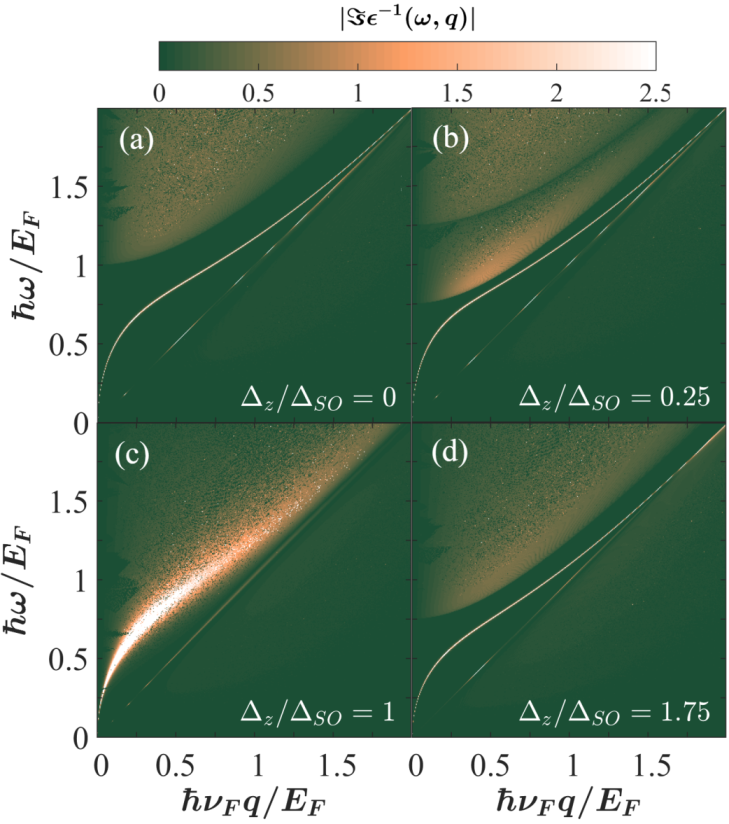}
	\caption{Loss function of DLS for $ \Delta_{SO}/E_F=0.5 $ and varying $ \Delta_z/\Delta_{SO}:\;(a)\,0,\, (b)\,0.25,\, (c)\,1$ and $(d)\,1.75 $ at $ T/T_F=0.5$.}
	\label{LossFunction}
\end{figure}

In order to study the behavior of the finite-temperature plasmonic dispersion in a DLS, we numerically calculate and plot the temperature-dependent DLS loss function, $\Im{(\epsilon^{-1}(\omega,q))} $, in the $ (\omega,q) $ space, for several parameters that affect the band structure. In Fig. \ref{LossFunction},  the loss function is plotted at $ T=0.5 T_F $ for the same parameters as in Fig. \ref{SPEs_0.5}. As this figure shows, the undamped acoustic and optical plasmon modes exist in the SPE gap for $ \Delta_z/\Delta_{SO}=0, 0.25, $ and $1.75 $  with the acoustic branches having very weak intensity. For $ \Delta_z/\Delta_{SO}=1 $, on the other hand, the acoustic and optical branches are broadened for most values of $ (\omega,q) $ displayed here. Since the SPE gap is almost closed and consequently, the plasmons are damped, there exist narrower peaks only for small $ q $'s indicating the undamped plasmons.

\begin{figure}[ht!]
	\vspace{5mm}
	\centering
	\includegraphics[width=8cm]{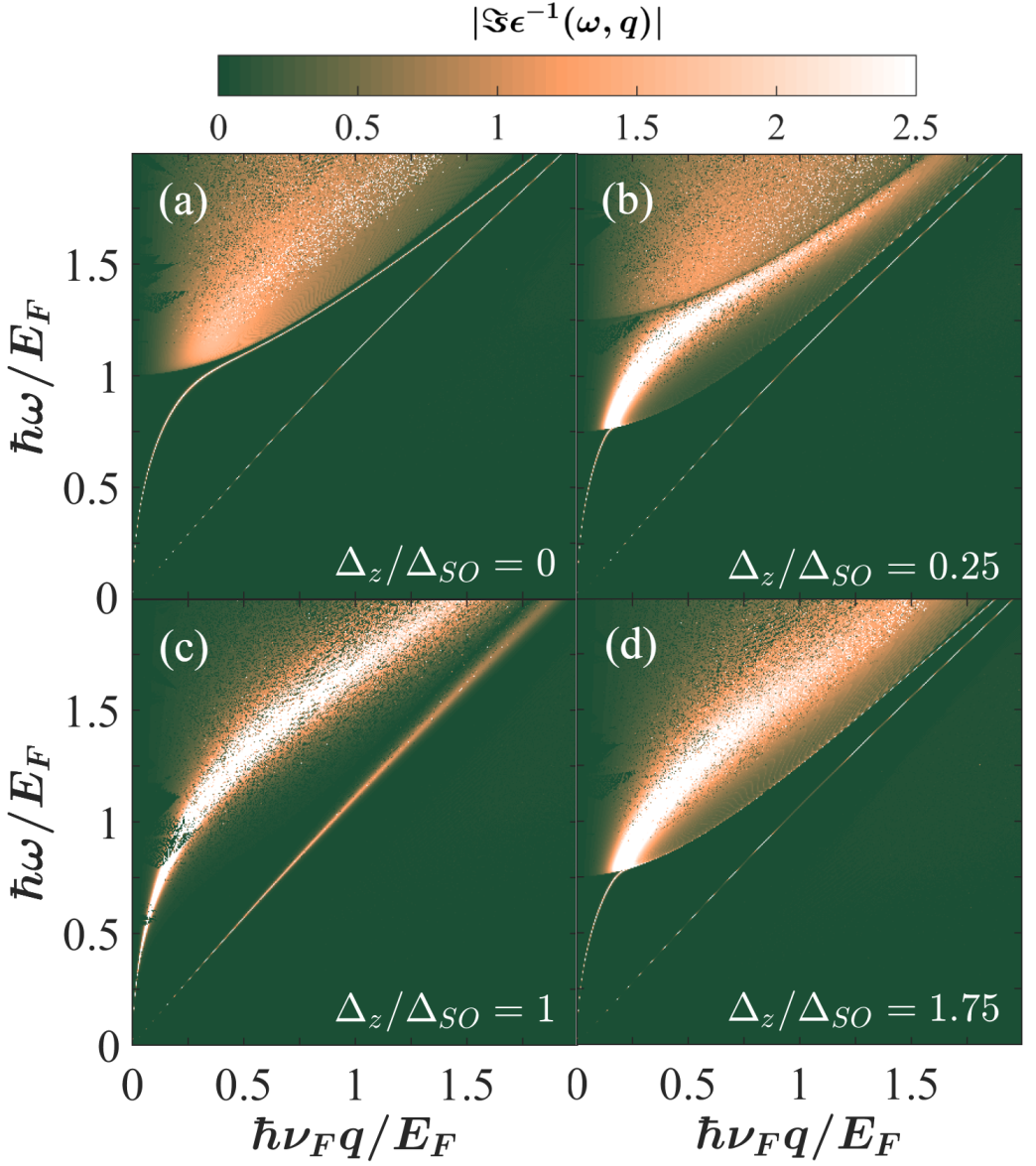}
	\caption{Loss function of DLS for $ \Delta_{SO}/E_F=0.5 $ and varying $ \Delta_z/\Delta_{SO}:(a)\,0,\, (b)\,0.25,\, (c)\,1$ and $(d)\,1.75 $ at $ T/T_F=1$.}
	\label{LossFunction_1}
\end{figure}

$\Im{(\epsilon^{-1}(\omega,q))} $ for the same parameters as Fig. \ref{LossFunction} but at $ T=T_F $ is depicted in Fig. \ref{LossFunction_1} where the emergence of new optical modes can be observed. When $ \Delta_z/\Delta_{SO}=0$ (panel \textcolor{blue}{(a)}), there is only one (undamped) optical branch. However, for $ \Delta_z/\Delta_{SO}=0.25 $ (panel \textcolor{blue}{(b)}) and $ \Delta_z/\Delta_{SO}=1.75 $ (panel \textcolor{blue}{(d)}) where both band gaps of silicene layers are nonzero, a new undamped optical branch is observed in the vicinity of the SPE$ ^{Inter}_{min} $'s boundary at higher values of $ q $. 
Moreover, a new damped mode appears in the weaker damping region of SPE continuum which is not extended from the undamped branch, continuously. Such splitting is not obtained for the acoustic branch. At $ \Delta_z/\Delta_{SO}=1 $ (panel \textcolor{blue}{(c)}), we see a similar trend for both branches as observed in Fig. \ref{LossFunction}\textcolor{blue}{(c)}. However, the branches are generally more broadened because the electron-hole pair excitations are facilitated by increasing temperature.

\begin{figure}[ht!]
	\centering
	\includegraphics[width=8cm]{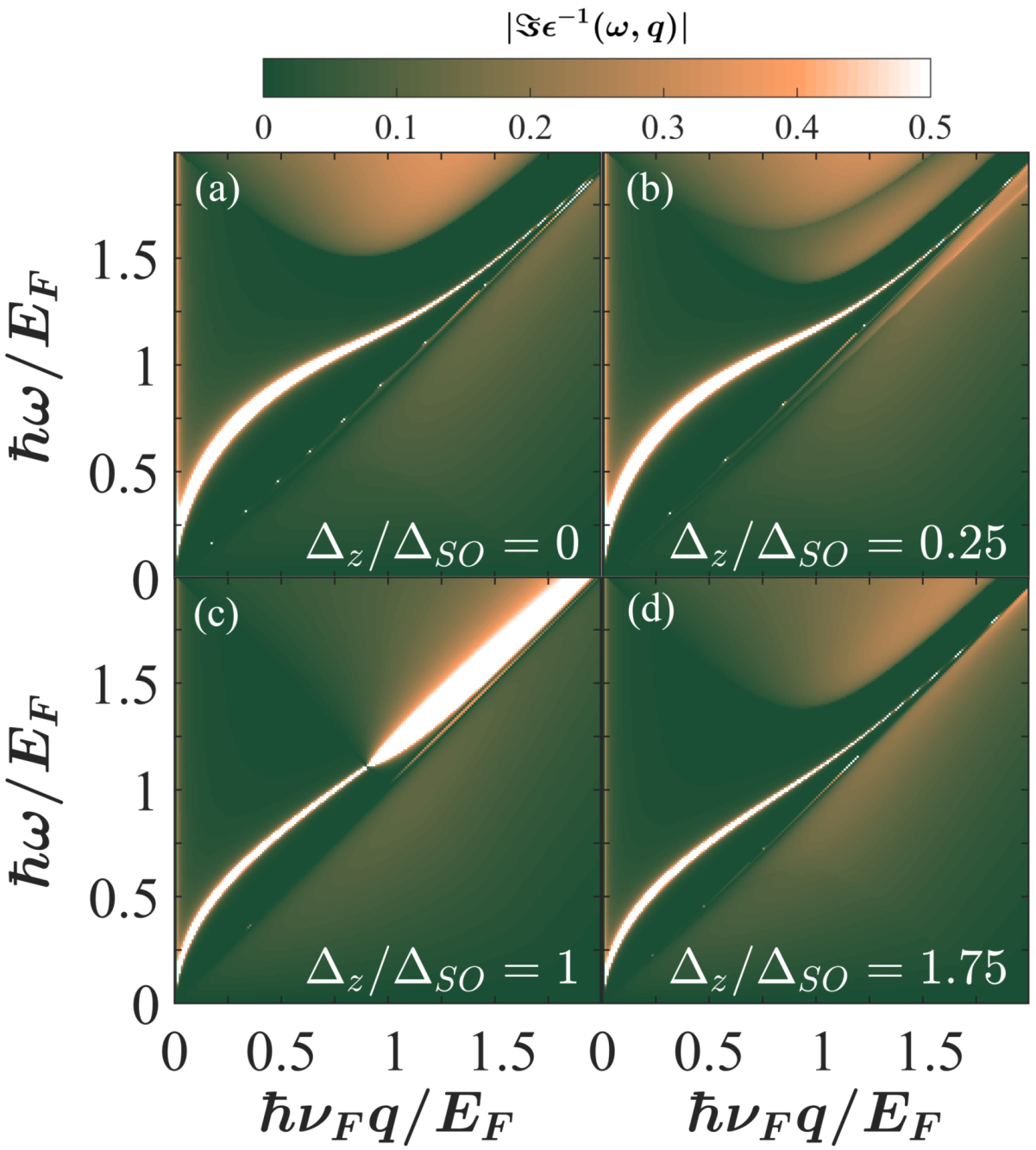}
	\caption{Loss function of DLS for $ \Delta_{SO}/E_F=0.5 $ and varying $ \Delta_z/\Delta_{SO}:(a)\,0,\, (b)\,0.25,\, (c)\,1$ and $(d)\,1.75 $ at $ T=0 $.}
	\label{LossFunction_Zero}
\end{figure}

Plots of the analytical zero-temperature loss function for the same parameters used for Fig. \ref{LossFunction} are shown in Fig. \ref{LossFunction_Zero}. The results resemble those in Fig. \ref{LossFunction} except for $ \Delta_z/\Delta_{SO}=1 $ (VSPM phase) where the undamped optical plasmons are present for a wider range of wave vectors.

\begin{figure}[htp]	
	\centering
	\includegraphics[width=8cm]{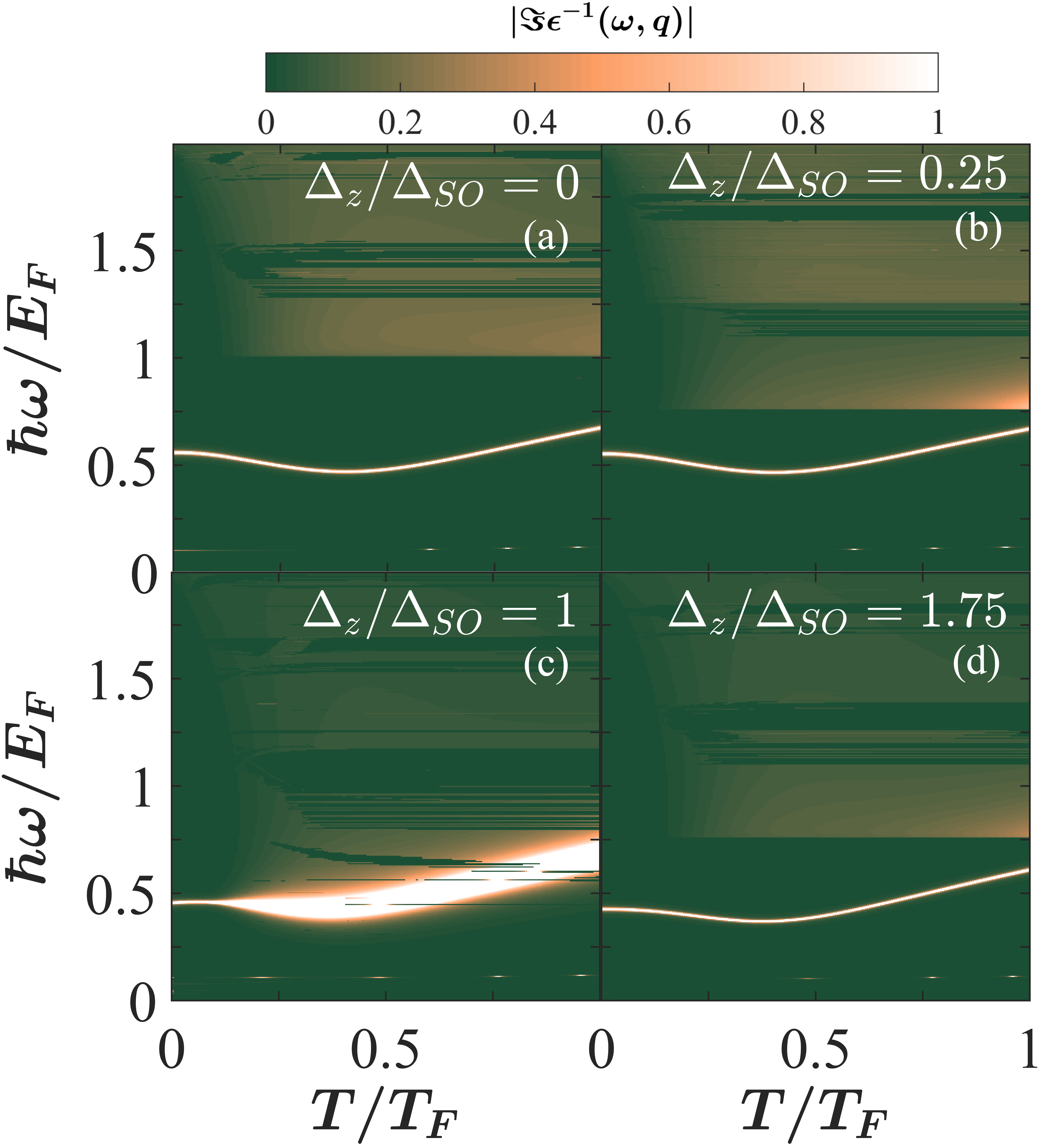}
	\caption{Loss function of DLS as a function of $ (\omega,T) $ for $ \Delta_{SO}/E_F=0.5 $ and two different values of $ \Delta_z/\Delta_{SO}:\;(a)\,0.25$ and $(b)\,1 $ at $ \tilde{q}=0.1 $.}
	\label{DoubleLossTemp}
\end{figure}

\begin{figure}[htp]
	\centering
	\includegraphics[width=8cm]{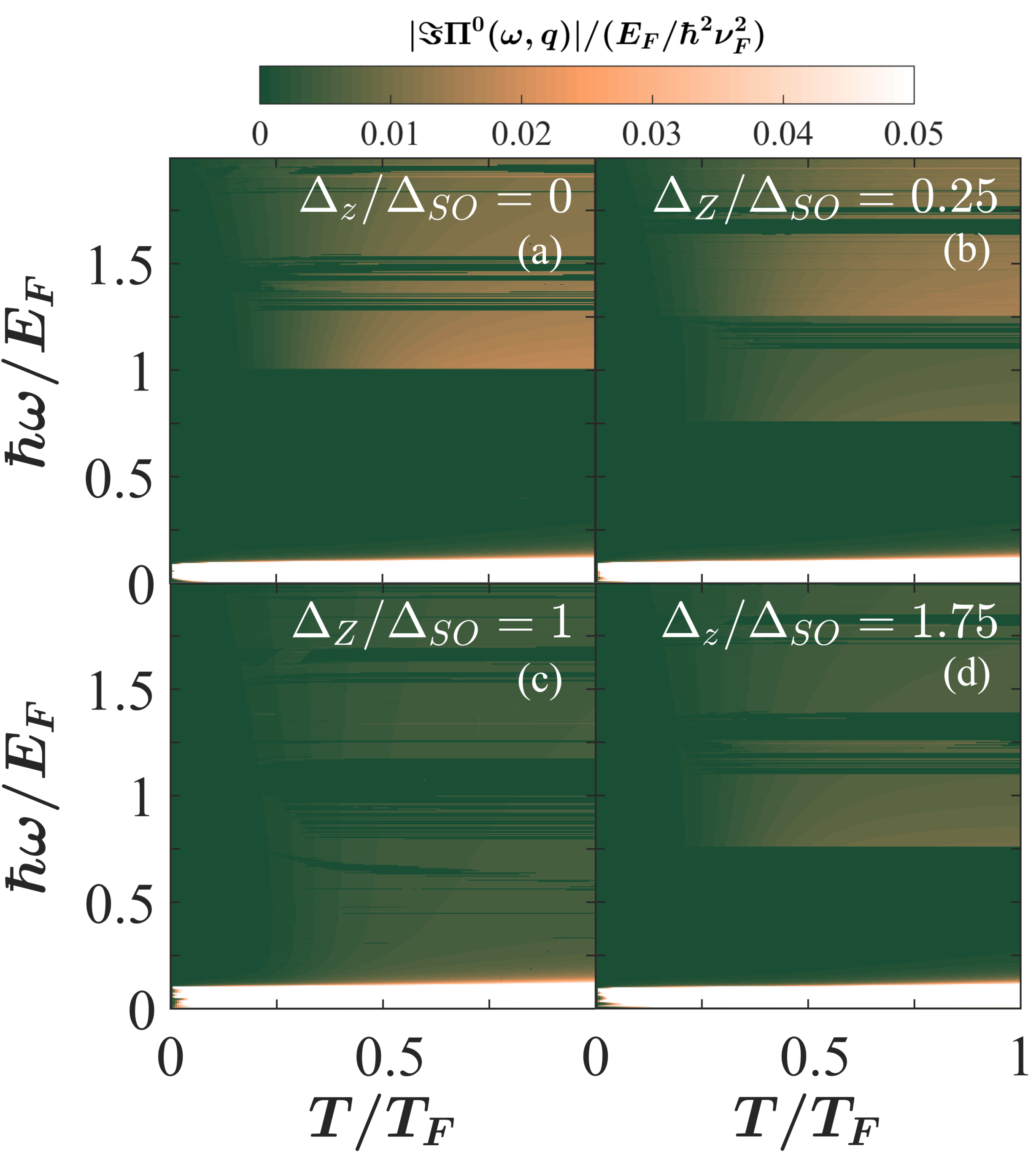}
	\caption{$ \Im{\Pi^0} $ as a function of $ (\omega,T) $, scaled by $ E_F/{{(\hbar\nu_F)}^2} $ for $ \Delta_{SO}/E_F=0.5 $ and two different values of $ \Delta_z/\Delta_{SO}:\;(a)\,0.25$ and $(b)\,1 $ at $ \tilde{q}=0.1 $.}
	\label{SPE_Temp_combined}
\end{figure}

Fig. \ref{DoubleLossTemp} illustrates the DLS loss function, $\Im(\epsilon^{-1}) $, over the $ (\omega,T) $ plane, at a given $ \tilde{q}=0.1 $, for different valuse of $ \Delta_z/{\Delta_{SO}}=0, 0.25, 1 $  and $1.75 $ with $ \Delta_{SO}/{E_F}=0.5 $. The upper and lower pronounced branches correspond to the optical and acoustic modes, respectively. Fig. \ref{SPE_Temp_combined} displays the extent of the SPE boundaries for the same parameters. As it can be observed from these figures, the optical branches are positioned in the SPE gap and, therefore, remain undamped for all presented values of $ T $, except for the VSPM phase. Furthermore, the frequency of the optical branch decreases with increasing temperature and reaches a minimum value around $ T=0.4T_F $ but after that it bounces off at higher temperatures. The behavior of optical plasmons for this studied DLS is similar to the plasmon dispersion of SLGG where the plasmon frequencies follow first a decreasing and then an increasing trend with a minimum at $ T \sim 0.5T_F $ for all band gap values \cite{patel2015} and similar to the double-layer graphene at small wave vectors\cite{tuan2013}. Recently, it has been shown that the temperature-dependent plasmons of SLS, SLG and single-layer n-doped and p-doped molybdenum disulfide can also exhibit such a behavior. \cite{iurov2016}

In Fig. \ref{DoubleLossTemp}\textcolor{blue}{(c)}, for $\Delta_z/\Delta_{SO}=1$, the optical branch follows a similar temperature trend as the other phases, but Fig. \ref{SPE_Temp_combined}\textcolor{blue}{(c)} shows that there is almost no gap left for the undamped plamons to reside. However, for $ T=0 $ and low temperatures, SPE$^ {Inter} $ boundary sits higher than the approximate interband SPE boundary. Thus, as can be seen, for $ T/{T_F}<0.1 $, the optical branch has a well-defined delta function-like peak which is a characteristic of undamped plasmons. Beyond this temperature, however, the peak is extremely broadened which indicates damping of plasmons at higher temperatures. Moreover, Fig. \ref{DoubleLossTemp} demonstrates the fact that the spectral weight of acoustic mode is insignificant compared to that of optical mode; this is an obstacle in experimentally observing the acoustic mode.
Applying a more distinguishing color limit to all panels of Fig. \ref{DoubleLossTemp}, reveals that the acoustic branches located in close proximity of SPE$^{Intra} $ boundary, grow negligibly with temperature which may not be of experimental significance. This figure also displays that as temperature rises, the interband boundary changes a bit where the intraband boundary moves slightly to higher frequencies.
\begin{figure}[ht]
	\centering
	\includegraphics[width=8cm]{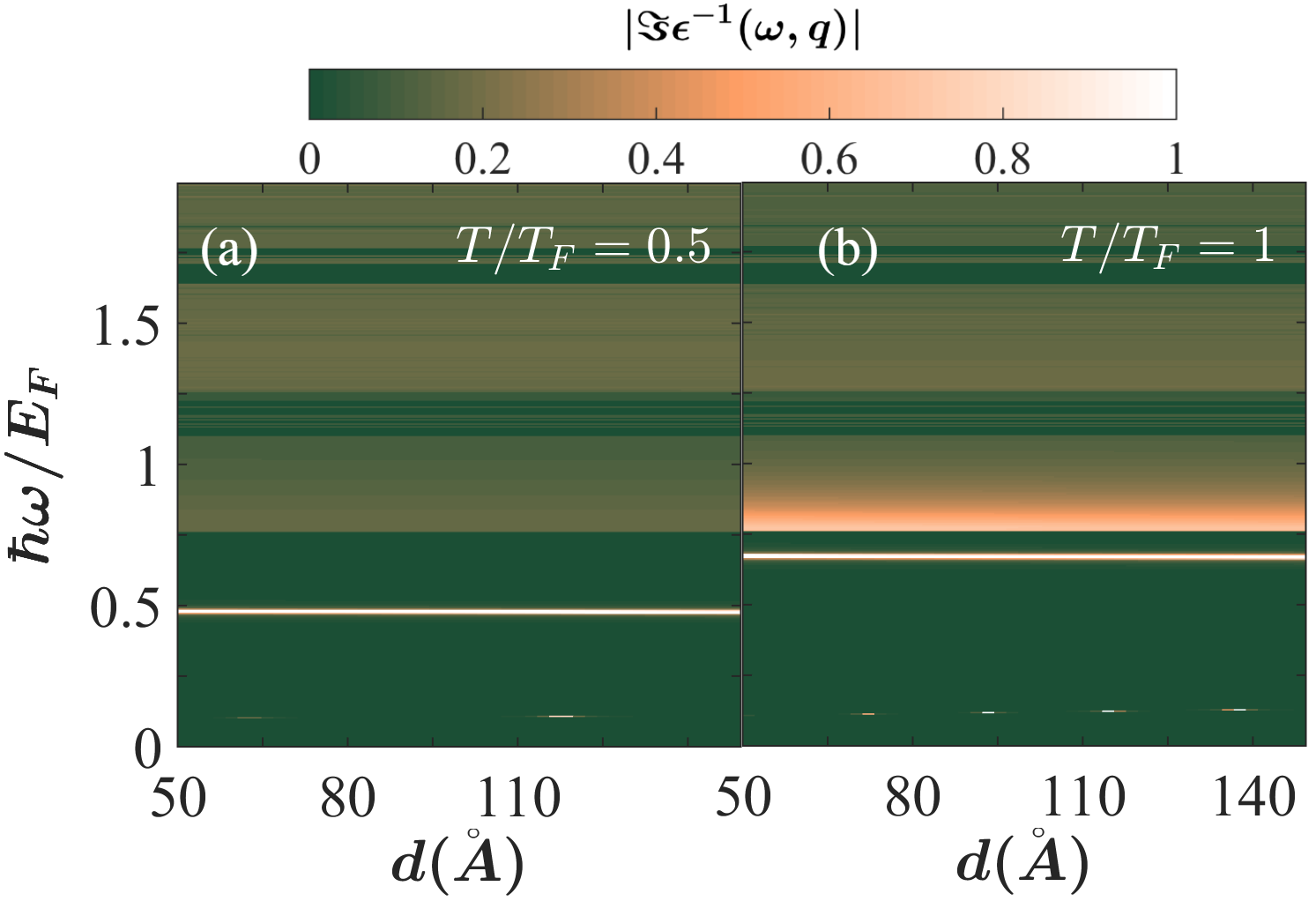}
	\caption{Loss function of DLS as a function of $ (\omega,d) $ at $ \tilde{q}=0.1 $, for $ \Delta_{SO}/E_F=0.5 $ and $ \Delta_z/\Delta_{SO}=0.25 $ at two different values of $ T/T_F:\;(a)\,0.5$ and $(b)\,1$.}
	\label{DoubleLossd}
\end{figure}

To investigate the effect of spatial separation between two layers on the DLS plasma oscillation modes, we plot $ \Im\epsilon^{-1}$ in the $ (\omega,d) $ plane at $ \tilde{q}=0.1 $ for $ \Delta_{SO}/{E_F}=0.5 $ and $ \Delta_z/{\Delta_{SO}}=0.25 $ in Fig. \ref{DoubleLossd} at two different temperatures, $ T=0.5T_F $ and $ T_F $. The interlayer separation varies from \unit[50]{\AA} up to \unit[150]{\AA} in this figure.
As demonstrated, for all the presented separation values at both temperatures, the undamped acoustic and optical modes exist (with a very weak acoustic mode spectral weight). Fig. \ref{DoubleLossd} suggests that, the finite-temperature optical modes are not affected by changing the spatial separation between the layers, similar to the zero-temperature case. On the other hand, applying a more distinguishing colormap displays that the frequency of acoustic plasmons increases slightly with $ d $  for both temperatures such that $ T=T_F $ dispersion has a slightly higher slope. This growth, is also observed at $ T=0 $ which is evident from the d-dependence of Eq. (\ref{ApproxBrach-}). Our results here may not be easily observed in experiment, though.

\begin{figure}[htp]
	\centering
	\includegraphics[width=8cm]{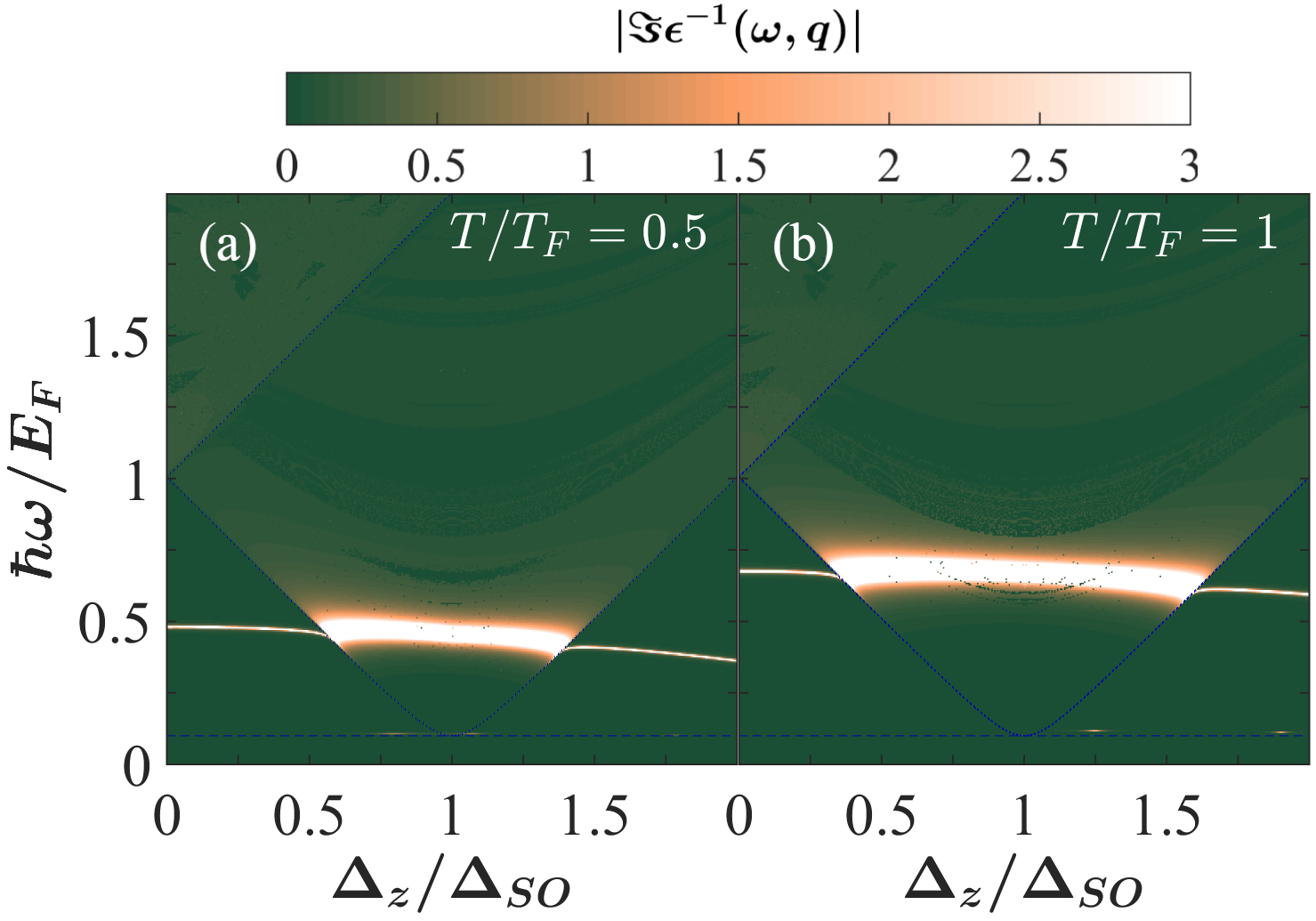}
	\caption{Loss function of DLS as a function of $ (\omega,\Delta_z) $ at $ \tilde{q}=0.1 $ for $ \Delta_{SO}/E_F=0.5 $ at two different values of $ T/T_F:\;(a)\,0.5$ and $(b)\,1$.}
	\label{DoubleLossdeltaZ}
\end{figure}

In order to study the effect of an external electric field on the finite-temperature DLS plasmons, the loss function is plotted in Fig. \ref{DoubleLossdeltaZ} for $ \Delta_{SO}/E_F=0.5 $ with $ T=0.5T_F$ and $ T_F $ at $ \tilde{q}=0.1 $ where dotted blue curves determine the SPE boundaries of Eq. (\ref{Boundary}). At both temperatures, the optical plasmon frequency decreases as $ \Delta_z $ increases quite similar to the trend for $ T=0 $ (see Fig. \ref{zeroPlasmon_DeltaZ}). Moreover, the optical plasmon branch enters the SPE region for a certain interval of $ \Delta_z/\Delta_{SO} $ around $ \Delta_z/\Delta_{SO}=1 $ and this interval is extended as the temperature increases from $ T=0.5 T_F $ up to $ T=T_F $. It can also be seen that the damped part of the optical plasmon branch is not extended continuously from the undamped part and it really seems more like a separate mode; for the acoustic branch however, such discontinuity can not be observed as it enters and exits the SPE region.
\section{Conclusion}
\label{sec:final}
To summarize, we have studied the finite-temperature plasmons of an n-doped double-layer 2D system consisting of silicene sheets with $ \Delta_{SO}=\unit[3.9]{meV} $ and $ \Delta_{SO}/E_F=0.5 $. We have considered all three phases of silicene (TI, VSPM and BI) by applying an external electric field to the system. For comparison, the zero- temperature results have been presented. By calculating the poles of the loss function, both the optical and acoustic plasmon branches have been obtained. To illustrate the finite-temperature SPE regions, we have plotted the imaginary part of the polarization function, as well. Our calculations for a DLS with $ d=\unit[100]{\AA} $ have shown that as the temperature is raised from $ 0 $ up to $ T_F $, the optical plasmon frequency first decreases and then increases smoothly for the different values of $ \Delta_z/\Delta_{SO}=0 $ (TI),$ 0.25$ (TI),$ 1$ (VSPM) and $1.75$ (BI). However, the acoustic modes change negligibly with temperature. Moreover, for a DLS with $ \Delta_z/\Delta_{SO}=0.25 $ at finite temperature, similar to the zero-temperature analytical results at long wavelengths, the optical plasmon dispersion is unaffected by changing the layer separation, at both $ T=0.5T_F $ and $ T=T_F $, while the acoustic plasmon dispersion exhibits a small growth. The effect of external electric field has also been studied for $ T=0.5T_F$ and $T_F $ at a given $ q $ and a decreasing trend has been observed for the optical plasmon dispersion as $ \Delta_z/\Delta_{SO} $ varies from $ 0 $ to $ 2 $ and a Landau damping behavior obtained around $ \Delta_z/\Delta_{SO}=1 $. The corresponding zero-temperature results have displayed a discontinuity at VSPM phase for both optical and acoustic plasmons.   
%
\end{document}